\documentclass[aip,jcp,a4paper,preprint]{revtex4}

\usepackage{placeins}
\usepackage{graphicx}
\usepackage{dcolumn}
\usepackage{bm}
\usepackage[]{placeins}
\usepackage{subfigure}
\usepackage{amsmath}
\usepackage[usenames,dvipsnames,svgnames,table]{xcolor}

\newcommand{\G}{\Gamma}
\newcommand{\w}{\omega}
\newcommand{\e}{\mathrm{e}}
\newcommand{\eps}{\varepsilon}
\newcommand{\g}{\gamma_{\scriptscriptstyle +}}
\newcommand{\gq}{\gamma_{\scriptscriptstyle -}}

\begin{document}
\title{The frequency-dependence of nonlinear conductivity in disordered systems: an analytically solvable model}
\date{\today}
\author{Clara Mattner}
\email{claramattner@uni-muenster.de} \affiliation{Westf\"alische
Institute for Physical Chemistry, WWU M\"unster,  Germany  }
\author{Bernhard Roling}
\email{roling@staff.uni-muenster.de} \affiliation{Department of
Chemistry, University of Marburg, Germany}
\author{Andreas Heuer}
\email{andheuer@uni-muenster.de} \affiliation{Institute for
Physical Chemistry, WWU M\"unster,  Germany and Center for
Nonlinear Science (CeNoS), WWU M\"unster,  Germany }

\begin{abstract}
For the hopping dynamics in a one-dimensional model, containing
energy and barrier disorder,  we determine the linear and
nonlinear response to an external field for arbitrary external
frequencies. The calculation is performed in analytical terms. We
systematically analyze the parameter space and find three
different regimes, corresponding to qualitatively different
frequency dependencies of the nonlinear response. Two regimes
agree with the results of recent conductivity experiments on
inorganic ion conductors and ionic liquids, respectively. The
ratio of the nonlinear and linear conductivity in the dc-regime
can be explicitly expressed  in terms of the disorder parameters.
As a generic feature the nonlinear conductivity displays a minimum
as a function of frequency which can be identified with
forward-backward dynamics in a double-well potential. The magnitude and sign of
the nonlinear conductivity around the minimum is a measure of the
disorder, inherent in this model.  Surprisingly, the frequency of
the minimum is hardly influenced by the disorder.
\end{abstract}

\maketitle

\section{Introduction}

Nonlinear ion transport in disordered systems, like glasses and
liquids, is of interest for mainly two reasons: (i) In
electrochemical devices, the integration of thin-film electrolytes
reduces the overall electrical resistance. In thin films, even
small voltage drops may lead to high electric fields and to a
field-dependent ionic conductivity. (ii) From a basic science
point of view, the study of nonlinear effects yields additional
information about ion transport mechanisms.

When high dc electric fields $E_{dc}$  {are} applied to isotropic
ionic conductors, the  {field dependence of the} nonlinear current
density
$j_{dc}(E_{dc})$ {can} be described in a first approximation by 
\cite{Vermeer56, Zagar69, Lacharme78, Hyde86, Barton96, Isard96}:
\begin{equation}
\label{eq_sinh} j_{dc} = j_0 \sinh \left( \frac{q\,a_{\rm
app}\,E_{dc}}{2\,k_B T} \right)
\end{equation}
 Here, $q$ denotes the ionic charge and $a_{\rm app}$ the 'apparent jump
distance'.

A experimental method for measuring nonlinear conductivities  is
the application of large ac electric fields $E(t) = E_0 \cos
(\omega\,t)$. In the dc-limit $E_0$ can be identified with
$E_{dc}$. Nonlinear conductivity {then leads to} the presence of
higher {order harmonic terms} for the resulting current. For
isotropic systems {nonlinear effects are reflected in terms
proportional to $\cos(3 \omega t), \sin(3 \omega t)$ and higher
harmonics}. Taking into account only
 {those terms which represent conductivity} in phase with the
electric field  one can generally write \cite{Roling08}:
\begin{equation}
\label{jfirst}
 j(t) = \left (\sigma_1^\prime(\omega) + O(E_0^2) \right ) E_0\cos(\omega t) +  {\tfrac{1}{4}} \left (\sigma_3^\prime(\omega) + O(E_0^2) \right ) E_0^3
\cos(3 \omega t).
\end{equation}

In general $\sigma_n^\prime(\omega)$ is frequency dependent for
disordered systems. For the dc-limit the validity of
Eq.\ref{eq_sinh} translates into the following expressions:
$\sigma_1^\prime(\omega \rightarrow 0) = j_0 q a_{\rm app}/(2k_B T)$
and $\sigma_3^\prime (\omega \rightarrow 0) = j_0 \bigl(q a_{\rm
app}/(2k_B T)\bigr)^3/6$.

 Via the relation
\begin{equation}
\label{eq_ratio_sig3_sig1} \frac{\sigma_{3}^\prime(\omega
\rightarrow 0)}{\sigma_{1}^\prime(\omega \rightarrow 0)} =
\frac{1}{6} \left( \frac{q\,a_{\rm app}}{2\,k_B T}\right) ^2
\end{equation}
it is possible to determine the apparent hopping distance $a_{\rm
app}$ from knowledge of  {$\sigma_{3}^\prime(\omega \rightarrow
0)$ and $\sigma_{1}^\prime(\omega \rightarrow 0)$}.

Typically, measured values for $a_{\rm app}$ range between 1.5~nm
$\;$ and 3~nm $\,$ \cite{Vermeer56, Zagar69, Lacharme78, Hyde86,
Barton96, Isard96}. These values are much larger than the typical
hopping distance $a_{\rm hopp}$ of ion transport which, e.g., for
alkali silicate systems is around to 0.25~nm \cite{Banhatti}. It
has been shown theoretically that{,} in a disordered potential
landscape, the value of $a_{\rm app}$ cannot be related to $a_{\rm
hopp}$  {in a simple way} \cite{Roling08}. Only in a regular
potential, $a_{\rm app}$ is identical to the hopping distance
$a_{\rm hopp}$ of the ions.

Experimentally, nonlinear ac measurements have been carried out on
alkali-ion conducting glasses as well as on the ionic liquid
HMIM-TFSI \cite{Roling08, Murugavel05, Heuer05, Staesche10,
Staesche10a, Feiten12}. In both cases, the nonlinear conductivity
$\sigma_3^\prime$ was positive in the  {dc regime} and gradually
decreased with increasing frequency. In the case of the alkali ion
conducting glasses, $\sigma_3^\prime$ became negative in the
{dispersive regime}. Furthermore, the onset frequency for the
linear and nonlinear conductivity, from which on deviations from
the dc-regime become relevant, are nearly identical. In contrast,
$\sigma_3^\prime$ remained positive for the ionic liquid {over the
entire frequency range}. Furthermore it turned out that the
frequency-dependence already starts at lower frequencies for the
nonlinear conductivity \cite{Feiten12}. Finally we mention that
from general arguments it follows (see, e.g. \cite{Roling08}) that
one expects a positive high-frequency regime. This has been
explicitly shown, e.g.,  for simulations of a one-dimensional
hopping model \cite{Roling08}.

The theoretical understanding of the ion dynamics is complex
due to the strong interaction among the mobile entities
\cite{ingram:1987,dyre:2009,maass:1999,Funkerecent}. Several
 groups have studied single-particle hopping motion in a discrete
disordered energy landscape to learn about the linear conductivity; see, e.g.,
\cite{hunt:1991,baranovskii:1999,dyre:2000,dyrenew}. Recently, this
approach has found a numerical justification since to a good
approximation the ion dynamics can be mapped on a single-particle
vacancy dynamics between distinct sites \cite{lammert:2010}.

For 1D models it is  possible to analytically calculate the
linear and nonlinear dc-current \cite{Heuer05,Kehr97,Eimax}.
Interestingly, for periodic boundary conditions the conductivity
displays non-analytic behavior at zero field in the thermodynamic
limit \cite{Roling08,Eimax}.
From studying a
disordered discrete 3D-energy landscape via computer simulations
it turned out that for a Gaussian distribution of site energies
the experimental relation $a_{\rm app} \gg a_{\rm hopp}$ can be
recovered  \cite{Rothel}. Surprisingly, for box-type distribution
$\sigma_3^\prime(\omega \rightarrow 0)$ is even negative{,} so
that according to Eq.\ref{eq_ratio_sig3_sig1} it {is} not possible
to get a real value for $a_{\rm hopp}$. Thus, a priori the
observation $a_{\rm app} \gg a_{\rm hopp}$ is non-trivial.

To  {understand} the frequency-dependence of
$\sigma_3^\prime(\omega)${,} the dynamics within a double-well
potential {were} studied in \cite{Roling2002}. Except for the
positive high-frequency plateau, $\sigma_3^\prime(\omega)$ is
negative throughout the remaining frequency range.  {In} general
it is not evident whether the restriction to a double-well
potential is sufficient to explain the frequency-dependence of
$\sigma_3^\prime(\omega)$.

Here we present a periodic 1D model which,  on the one hand, displays
typical behavior of disordered systems but, on the other hand, is
simple enough so that the frequency dependence of the linear and
the nonlinear conductivity can be calculated analytically.
Discussion of this model will  {broaden the understanding of} the
experimental observations described above. This model contains
three parameters, reflecting the typical barrier heights, the site
disorder and the barrier disorder. Two questions are of particular
relevance: (1) How do these parameters influence the qualitative
frequency-dependence of the nonlinear conductivity? (2)  To which
degree can the theoretical description of the minimum in the
nonlinear conductivity be reduced to the dynamics in local
double-well potentials?

The paper is organized as follows. In  {Sect.~II} we introduce the
model. The analytical results can be found in {Sect.~III}. They
will be discussed in  {Sect.~IV}. We conclude in
 {Sect.~V}.

\section{Model}

\begin{figure}[h!]
    \centering
   \includegraphics[width=0.60\textwidth]{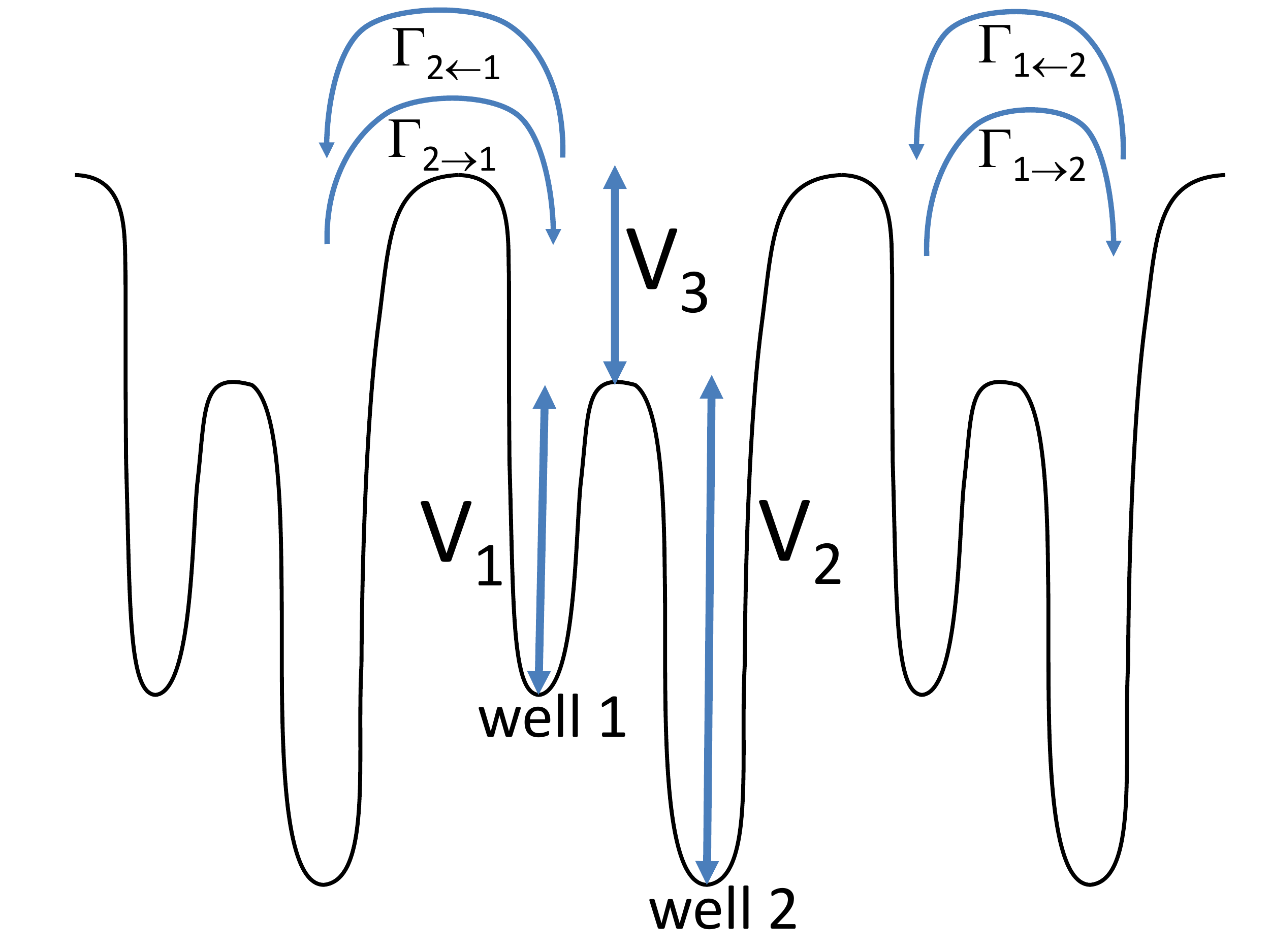}
   \caption{\label{fig:unendlich} Sketch of the hopping model,
   used in this analysis.   
   }
\end{figure}

As a model we consider a periodic potential with two different
wells. It is characterized by the three energy values $V_1$, $V_2$
and $V_3$; see Fig.\ref{fig:unendlich}. In what follows we always
consider $V_1 \le V_2$. A particle performs activated hopping
between adjacent sites. We introduce the probabilities
$y_{1,2}(t)$ that the particle is in the left or the right well,
respectively. In particular one has $y_1(t) + y_2(t) = 1$. The
activation barriers are modulated by the external potential
$V_{ext}(t) = \eps\cos\w t$.

In case of a regular potential, i.e. $V_1 = V_2$ and $V_3=0$ this
simple model yields
\begin{equation}
j_{dc} \propto \sinh \left(\frac{\varepsilon}{k_BT}\right).
\end{equation}
Via comparison with Eq.\ref{eq_sinh} we can relate $\varepsilon$
to the experimental parameters, i.e.
\begin{equation}
\label{identify}
 \varepsilon = q a_{\rm app} E_0/2.
\end{equation}

As mentioned above, for this regular potential {one has} $a_{\rm
app} = a_{\rm hopp}$. For reasons of simplicity we will express
all our results in terms of $\varepsilon$. Furthermore, we choose
$k_BT {=1}$, i.e. we express all energies in terms of $k_BT$.

There are two possible transitions from well 1 to 2 and vice
versa. This yields for the transition rate from well 1 to well 2
\begin{equation}
 \G_{12} = \G_{1\rightarrow 2} +\G_{2\leftarrow 1}
 \end{equation}
 with
 \begin{equation}
\G_{2\leftarrow 1}  = \Gamma_0 \exp(-(V_1+V_3)-\varepsilon
\cos(\omega t))
 \end{equation}
 and
  \begin{equation}
\G_{1\rightarrow 2}  = \Gamma_0 \exp(-V_2 + \varepsilon
\cos(\omega t))
 \end{equation}
For  {simplification} we choose the kinetic prefactor $\Gamma_0$
to be unity.  Introducing the notation  $\Gamma_i:=\e^{-{V_i}}$
one can therefore write
\begin{equation}
      \G_{12}=  \G_1\e^{\eps\cos\w t}+\G_1\G_3 \e^{-\eps\cos\w t} \label{g12}.
\end{equation}
In analogy one obtains
\begin{equation}
  \G_{21} = \G_2\e^{-\eps\cos\w t} + \G_2\G_3\e^{\eps\cos\w t} \label{g21}.
 \end{equation}

Now one can write the rate equation for $y_1(t)$ as

\begin{align}
  \dot{{y}}_1(t) &=-\G_{12}(t){y}_1(t)+\G_{21}(t)(1-{y}_1(t))  \\
  &= - {y}_1(t) \left(( \G_1+ \G_2\G_3)\e^{\eps\cos\w t} + ( \G_2+ \G_1\G_3)\e^{-\eps\cos\w t}\right) \nonumber \\
             &\qquad + \G_2\e^{-\eps\cos\w t} +  \G_2\G_3\e^{\eps\cos\w t} \label{y1dgl}
\end{align}

To simplify this rate equation we introduce the equilibrium population of $y_1$, denoted
   as $p_1$, given by the Boltzmann factor ($\Delta V = V_2 - V_1$) 
   \begin{equation}
p_1:=\frac{\e^{-\Delta V/2}}{\e^{\Delta V/2}+{\e^{-\Delta V/2}}} \label{a}
\end{equation}
and  {define} $u_1(t)$ as the difference to the equilibrium
population, i.e.
\begin{equation}
u_1(t)   := y_1(t) - p_1. \label{udef}
\end{equation}
Then a straightforward calculation \cite{Bachelor} yields from
Eq.\ref{y1dgl}
\begin{equation}
 \dot{ {u}}_1(t) = - {u}_1(t)\Bigl( \Gamma_a\e^{\eps\cos\w t}+\Gamma_b\e^{-\eps\cos\w t} \Bigr) +  {\Gamma}\Bigl(-\e^{\eps\cos\w t}+\e^{-\eps\cos\w t}\Bigr). \label{DGl}
\end{equation}
Here we  {used} the abbreviations \label{gammaa}$\G_a:= \G_1+
\G_2\G_3$  {and} $\G_b:= \G_2+ \G_1\G_3$  {as well as}
\begin{equation}
{\G}={\frac{-\e^{-(1/2)(V_1+V_2)-V_3}+\e^{-(1/2)(V_1+V_2)}}{\e^{\Delta V/2}+\e^{-\Delta V/2}}} \label{3stern}
\end{equation}

Note that after a transformation $V_1 \rightarrow V_1  - c$, $V_2
\rightarrow V_2  - c$ and $V_3 \rightarrow V_3$ with $c \le V_1$
and $c \le V_2$ the dynamics is the same except for a trivial
scaling of the hopping rate by $\exp(c)$. Thus, in the subsequent
analysis we can  {set}  $V_1 = 0$, i.e. $\Gamma_1 = 1$. Then,
$V_2$ can be interpreted as the asymmetry{,} $V_3$ as the barrier
disorder,  {Eqn.\ref{3stern} reduces to}
\begin{equation}
{\G} = \frac{\Gamma_2(1-\Gamma_3)}{1+\Gamma_2},
\end{equation}
 {and Eqn.\ref{a} to}
\begin{equation}
p_1 = \frac{\Gamma_2}{1+\Gamma_2}.
\end{equation}

\section{Analytical solution}
\subsection{Non-equilibrium population}

The general solution of $u_1(t)$ can be written as a sum over the different harmonics
$\cos(n \omega t)$ and $\sin(n \omega t)$, i.e.
\begin{equation}
u_1(t) = \sum_{n=0}^\infty a_n(\varepsilon) \cos(n \omega t) +
\sum_{n=1}^\infty b_n(\varepsilon) \sin(n \omega t). \label{ugen}
\end{equation}
Formally, the individual $a_n(\varepsilon)$ and $b_n(\varepsilon)$
can be written as a Taylor series in $\varepsilon$. One can easily
show that the lowest-order terms of $a_n(\varepsilon)$ and
$b_n(\varepsilon)$, respectively, are proportional to
$\varepsilon^n$.

To calculate the current which is in phase with the external field
only the cosine-terms, i.e. the $a_n(\varepsilon)$, are of
relevance. In what follows we define
\begin{equation}
\alpha_n := \lim_{\varepsilon \rightarrow 0}
(a_n(\varepsilon)/\varepsilon^n).
\end{equation}
Since for small fields $a_n \propto \varepsilon^n$, the $\alpha_n$
are independent of $\varepsilon$ and characterize the nonlinear
dynamics in the experimentally relevant case of weak nonlinear
effects. Similarly, one can define the $\beta_n$  {as new
coefficients for the sine terms.}

After inserting Eq.\ref{ugen} into Eq.\ref{DGl}, and expanding the
exponential terms in terms of the different harmonics ($\cos
(n\omega t)$, $\sin (n\omega t)$, obtained after application of
appropriate addition theorems) one can set all terms, belonging to
the same harmonics and the same exponent in $\varepsilon$, to
zero.  This yields a system of linear equations in the $\alpha_n$
and $\beta_n$ which can be solved iteratively, starting with
$n=1$. After a  tedious but straightforward calculation one ends
up with
\begin{equation}
\label{a1}
 \alpha_1=-\frac{2\g\G}{\w^2+\g^2},
\end{equation}
\begin{equation}
\label{a2}
{\alpha_2=\gq\frac{-\G(2\w^2-\g^2)}{(\w^2+\g^2)(4\w^2+\g^2)}}
 \end{equation}
 and
\begin{equation}
\label{a3}
 \alpha_3={\frac{\g\G(\g^2-5\w^2)}{6(9\w^2+\g^2)(\w^2+\g^2)}
 +\gq^2\frac{\g\G(11\w^2-\g^2)}{2(\w^2+\g^2)(4\w^2+\g^2)(9\w^2+\g^2)} }.
\end{equation}
Here we use the abbreviations
\begin{align}\label{gamma}
 \g&:=\G_a +\G_b = (1+\Gamma_2)(1+\Gamma_3)\\
 \gq&:=\G_a-\G_b = (1-\Gamma_2)(1-\Gamma_3). \label{gamma2}
\end{align}

\subsection{Calculation of the conductivity}

The dynamics of a charged particle in the periodic potential can be characterized
by a current. Based on the time-dependent populations as determined in the previous section one can generally write for the current
\begin{equation}
{j}(t)=\G_1\e^{\eps\cos\w t} {y}_1(t)- \G_2\e^{-\eps\cos\w t} {y}_2(t)+ \G_2\G_3\e^{\eps\cos\w t} {y}_2(t)- \G_1\G_3\e^{-\eps\cos\w t} {y}_1(t)\\
\end{equation}
which can be rewritten with $y_2(t) = 1 - y_1(t)$ as
\begin{equation}
{j}(t) = -\G_2\e^{-\eps\cos\w t}+\G_2\G_3\e^{\eps\cos\w t}+
{y}_1(t)\bigl(\G_c\e^{\eps\cos\w t}+\G_d\e^{-\eps\cos\w t}\bigr).
\end{equation}
Here we have used the abbreviations
$\G_c:=\G_1-\G_2\G_3$ and $\G_d:=\G_2-\G_1\G_3$\label{Gammac}.
From now on we will choose again $\Gamma_1 = 1$.

For the subsequent calculations the exponential-terms have to {be}
expanded with respect to $\varepsilon$. Furthermore, for the
population $y_1(t)$ the 
{expression from Eq.\ref{ugen} combined with Eq.\ref{udef}}
can be inserted. 
Then one needs to combine again all terms which scale with the
same harmonics and the same exponent in $\varepsilon$ in analogy
to the calculation of the $\alpha_n$. This results in an
expression for the current which contains terms proportional to
$\cos(n \omega t)$ {or $\sin(n \omega t)$} for all integer $n \ge
1${, as well as the coefficients $\alpha_n, \beta_n$}.

Macroscopic disordered systems behave isotropically, i.e. the
reversal of  {an} electric {field} leads to a reversal of the
current. In contrast, the present model is  anisotropic for $ V_2
\ne 0$. To enable a direct comparison with the experimental
situation{,} we therefore average over two opposite directions of
the field. As a consequence only the odd harmonics (i.e.
$\cos(\omega t), \cos(3 \omega t), ...$) and only terms which are
uneven with respect to $\varepsilon$ remain. All other terms
 {cancel}. Thus, one can finally write
\begin{equation}
j(t) = s_1(\varepsilon,\omega) \cos (\omega t) +
s_3(\varepsilon,\omega) \cos (3\omega t) +  {\text{sine terms,}}
\end{equation}
where the lowest order term of $s_n(\varepsilon,\omega)$ is
proportional to $\varepsilon^n$.  In analogy to the discussion of
the populations $\alpha_n$ we define the conductivities via
  \begin{equation}
\sigma_n(\omega) := \lim_{\varepsilon \rightarrow 0}
(s_n(\varepsilon,\omega)/\varepsilon^n),
\end{equation}
thereby capturing the effect of the external potential in lowest
order.

Via comparison with Eq.\ref{jfirst} one can identify $\sigma_1
\equiv \sigma_1^\prime$ and $\sigma_3  \equiv \sigma_3^\prime/4$.
The experimentally relevant quantity $a_{\rm app}/a_{\rm hopp}$
can thus be expressed as (also using Eq.\ref{identify})
\begin{equation}
\frac{a_{\rm app}}{a_{\rm hopp}} = \frac{24
\sigma_3(\omega=0)}{\sigma_1(\omega=0)}.
\end{equation}
More generally we define
\begin{equation}
A(\omega) = \frac{24 \sigma_3(\omega)}{\sigma_1(\omega)}.
\label{Ao}
\end{equation}

In what follows we give the results of  {the} lengthy but
straightforward calculations  {described above} \cite{Bachelor}.
{The in-phase terms} read
\begin{equation}
{\sigma}_1(\omega) = \frac{2 \Gamma_2 (1+\Gamma_3)}{1 + \Gamma_2}
+ (1+\Gamma_2)(1-\Gamma_3) \alpha_1 (\omega) \label{s1}
\end{equation}
and
\begin{eqnarray}
{\sigma}_3(\omega) &=& \frac{\Gamma_2 (1 +
\Gamma_3)}{12(1+\Gamma_2)}
+\tfrac{1}{8}{\alpha}_1 (\omega)(1+\Gamma_2)(1-\Gamma_3)+\tfrac{1}{2}{\alpha}_2 (\omega)( 1-\Gamma_2)(1+\Gamma_3) \nonumber \\
&+&{\alpha}_3 (\omega)( 1+\Gamma_2)(1-\Gamma_3). \label{s3}
\end{eqnarray}

These equations are the key result of this work because all
conclusions, discussed below, follow from them.

Of particular interest are the limits  {at vanishing or} infinite
frequency. After some algebra one obtains
\begin{equation}
{\sigma}_1(\omega=0) =  \frac{8 \Gamma_2 \Gamma_3}{
(1+\Gamma_2)(1+\Gamma_3)} , \end{equation}
\begin{equation}
{\sigma}_3(\omega=0) =  \frac{{\sigma}_1(\omega=0)}{24} \left(1 +
6
\frac{(1-\Gamma_2)^2}{(1+\Gamma_2)^2}\frac{(1-\Gamma_3)^2}{(1+\Gamma_3)^2}
\right ), \end{equation}
\begin{equation}
{\sigma}_1(\omega \rightarrow \infty) = \frac{2 \Gamma_2
(1+\Gamma_3)}{1 + \Gamma_2} ,
\end{equation}
and
\begin{equation}
{\sigma}_3(\omega \rightarrow \infty) = \frac{{\sigma}_1(\omega
\rightarrow \infty )}{24}. \label{s3i}
\end{equation}

Furthermore, it is also possible to express the whole
frequency-dependence of the linear conductivity by a simple
expression. It is given by
\begin{equation}
\sigma_1(\omega) = {\sigma}_1(\omega=0)\frac{\gamma_+^2 + \omega^2
\frac{(1+\Gamma_3)^2}{4\Gamma_3}}{\gamma_+^2 + \omega^2}
\label{sigmaall}
\end{equation}
Naturally, this expression is consistent with the two limiting
cases, given above.

\begin{figure}[h!]
    \centering
   \includegraphics[width=0.40\textwidth]{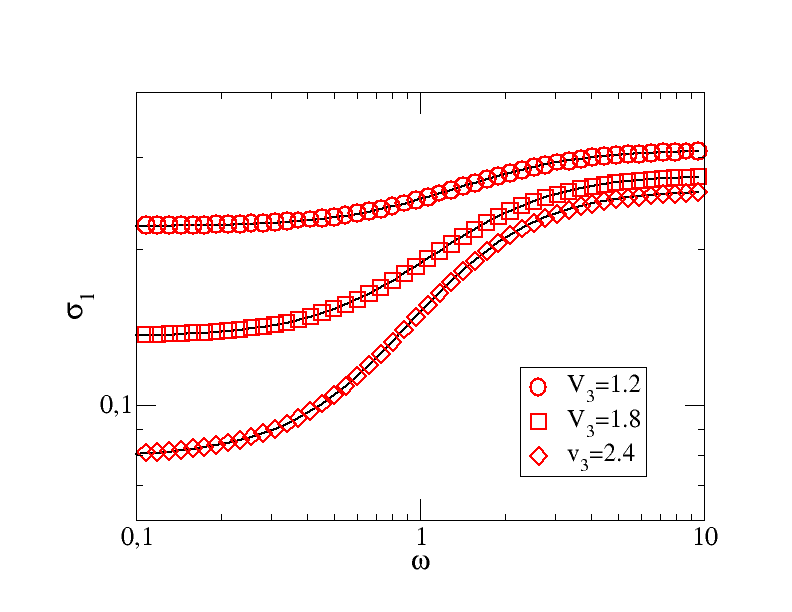}
      \includegraphics[width=0.40\textwidth]{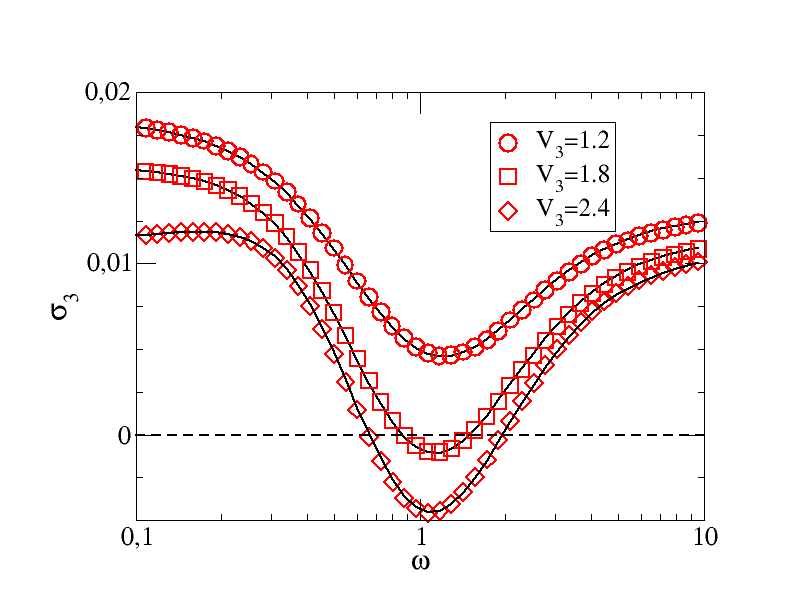}
   \caption{ Comparison of the numerical
   solution of the rate equations (symbols) with the analytical expressions  (lines)
   for $V_2 = 2$ and different values of $V_3$. Left: $\sigma_1(\omega)$, right:
   $\sigma_3(\omega)$. For \ {labeling} purposes  {in Fig.\ref{fig:phasediagram} the three parameter
   specifications are
  denoted a, b, c from top to bottom.}
   \label{anacomp}
   }
\end{figure}

Due to the complexity of the expressions we first check  {that
straightforward} numerical simulations of the corresponding rate
equations and subsequent determination of the current via Fourier
transformation agree with the analytical expressions. This is
shown in Fig.\ref{anacomp}. Indeed, one can find an excellent
agreement (also, for other values of $V_2$ and $V_3$  {not shown
here}) which strongly supports the correctness of our algebraic
calculations.

\section{Discussion}

{First, we discuss $A(\omega)$ as introduced in Eq.\ref{Ao}. For
the special case of an ordered potential (i.e. $V_2= V_3=0$), one
obtains $A(\omega)=1$. Generally,} in the high-frequency limit all
values $a_n$ disappear, i.e. the populations of the two wells do
not differ from the Boltzmann distribution. As a consequence,
{$A(\omega\rightarrow\infty)$} is unity. One can also see from
Eqs.\ref{a1} {-}\ref{a3} that for the specific case $\Gamma_3 = 1$
all $\alpha_n$  {disappear for} all finite frequencies.  As a
consequence, both the linear and the nonlinear conductivity are
frequency-independent and thus trivially display the same ratio
for all frequencies. This limit corresponds  to the case of
vanishing barrier disorder, i.e. $V_3 = 0$. A prototype model for
this scenario is the trap model \cite{Bouchaud}.

Of particular interest is the zero-frequency limit of $A(\omega)$.
Based on Eqs.\ref{s1} and \ref{s3} one obtains
\begin{equation}
A(\omega=0) = 1 + 6
\frac{(1-\Gamma_2)^2}{(1+\Gamma_2)^2}\frac{(1-\Gamma_3)^2}{(1+\Gamma_3)^2}.
\end{equation}
Thus, for $V_2=0$  {or} $V_3=0$ {, equivalently $\Gamma_2=1$ or
$\Gamma_3=1$, the proportion between the nonlinear and linear part
of the conductivity becomes minimal}. For large values of the
asymmetry and the barrier  {disorder,} $A(\omega=0)$ can approach
values as large as 7. Thus, for this model  {any} disorder
generally leads to $A(\omega=0) \gg 1$ in agreement with the
experimental observation. Interestingly, for the dc-behavior the
impact of the asymmetry is the same as the
impact of the barrier disorder. Naturally, if expressing these 
results in terms of temperature one would expect an increase of
$A(\omega=0)$ with decreasing temperature.

\begin{figure}
\centering
\includegraphics[width=0.5\linewidth]{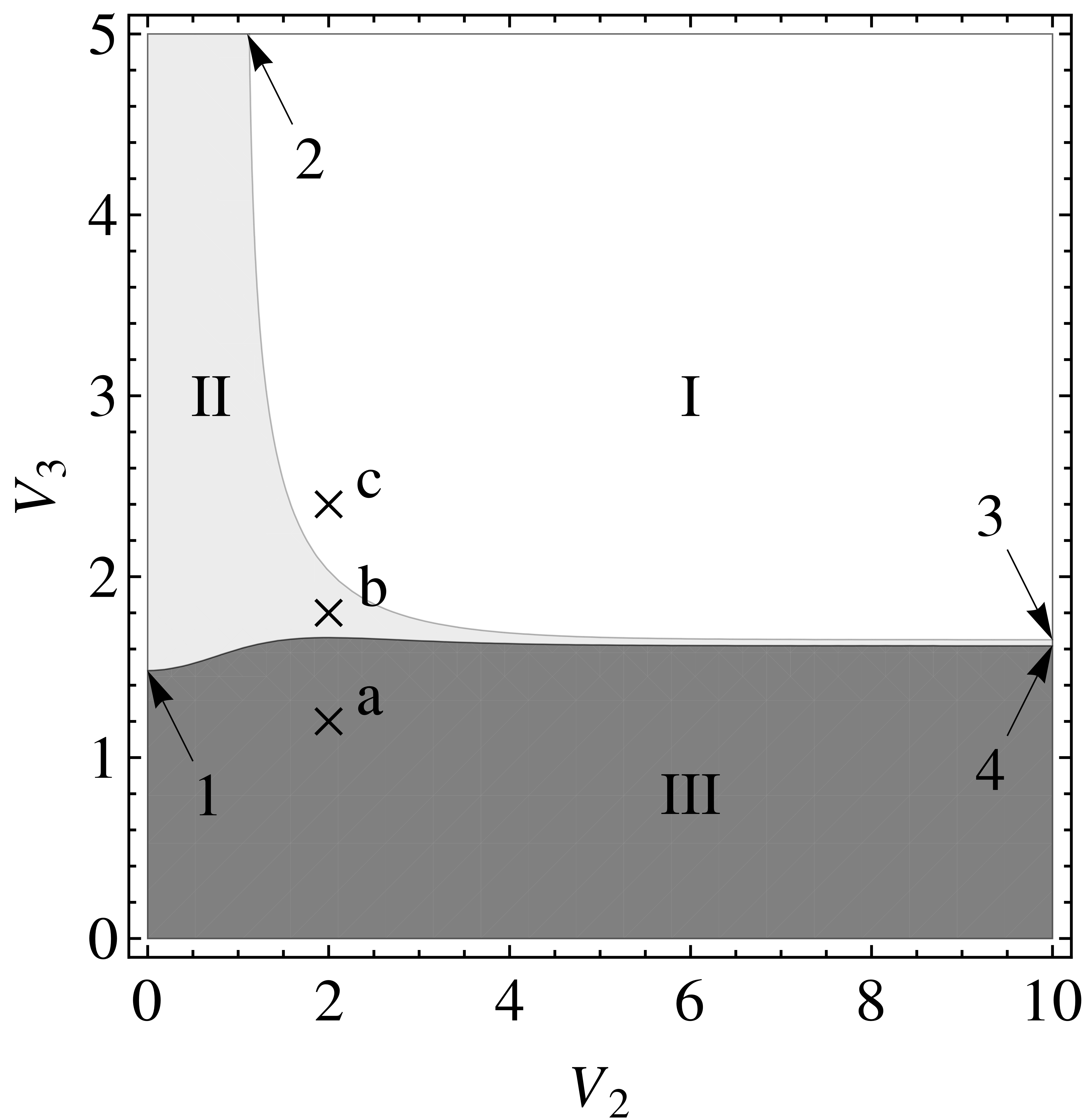}
\caption{Representation of the different regimes I, II, and III
(definition, see main text) concerning the qualitative behavior of
the frequency dependence of $\sigma_3(\omega)$. The letters
 {a,b,c mark the values $V_2$ and $V_3$ that are used for the
three curves in Fig.\ref{anacomp}.
The numbers 1 - 4 refer to the different limit cases} mentioned in the
text.} \label{fig:phasediagram}
\end{figure}

It turns out that in the whole parameter space $\sigma_3(\omega)$
displays one minimum as a function of frequency. However, as shown
in Fig.\ref{anacomp} {,} two other properties {, i.e. existence of
negative values  and sign of initial slope of $\sigma_3(\omega)$}
strongly depend on the specific values of $V_2$ and $V_3$ as shown
by variation of $V_3$ for fixed $V_2$ (here: $V_2$=2). First, for
 $V_3=1.2$ the nonlinear conductivity does not acquire
negative values whereas for $V_3 \ge 1.8$ negative values are
observed. Second, the low-frequency slope of $\sigma_3(\omega)$
is negative for $V_3 \le 1.8$ and positive else. {This shows that, for certain $V_2$, all 
possibilities can be realized through variation of $V_3$.} Via a
numerical analysis we have {analyzed} each parameter pair
$(V_2,V_3)$ with respect to these two properties. We have
identified three different regimes: (I) Positive initial slope and
presence of a negative frequency regime; (II) Negative initial
slope and presence of negative frequency regime; (III) Negative
initial slope and only positive values. The remaining possible
option does not occur. The experimental data for ion conductors
correspond to regime (II), those for ionic liquids to regime (I).
A systematic identification of all
three regimes can be found in Fig.\ref{fig:phasediagram}.

Interestingly, the presence  of a negative regime hardly depends
on the specific value of the asymmetry $V_2$ but rather on the
difference of the barrier heights $V_3$. In contrast, regime I is
approximately symmetric with respect to the values of $V_2$ and
$V_3$.

Some of the limiting values of this phase diagram can be predicted
{even} analytically. Due to the complexities of the resulting
algebraic equations we have employed an algebra program
(Mathematica). The following results  {were} obtained: (1) The
transition between regime II and III for  {$V_2=0$ occurs} for
$V_3 =  {\ln\bigl(\tfrac{1}{9}(29+4\sqrt{7})\bigr)} \approx 1.48$.
(2) The transition between regimes I and II for large $V_3$, i.e.
for vanishing slope of $\sigma_3(\omega)$, occurs for $V_2 = \ln 3
\approx 1.10$. (3) The transition between both regimes for large
$V_2$ occurs for $V_3 = \ln\bigl((73+10\sqrt{46})/27 {\bigr)
}\approx 1.65$. (4) The transition between regimes II and III for
large $V_2$ occurs for $V_3 \approx 1.62$. This  {last} value
emerges from solving a polynomial  {equation} of 12th order and
 {was} determined {only} numerically.

In the limit $V_3 \rightarrow \infty $, i.e. $\Gamma_3 = 0$ {,}
the periodic potential transforms into a double-well potential
{DWP} with vanishing dc-conductivity,  {so} $\sigma_1(\omega=0) =
\sigma_3(\omega=0) = 0$. In this case the system can either be
described by regime I or by regime II. In both cases
 {$\sigma_3(\omega)$ becomes negative for intermediate frequencies.}
Most importantly, upon variation of (mainly) $V_3$ the
nonlinearity does acquire negative values (large $V_3$ {, regime I
or II}) or only positive values (small $V_3$ {, regime III}).
Thus, variation of the sign of $\sigma_3(\omega)$ at its minimum
does not necessarily reflect different physical mechanisms. In
this sense the different experimental results for inorganic ion
conductors and ionic liquids  {mentioned in the Introduction} may
just reflect a different degree of disorder.

\begin{figure}
\centering
\includegraphics[width=0.45\linewidth]{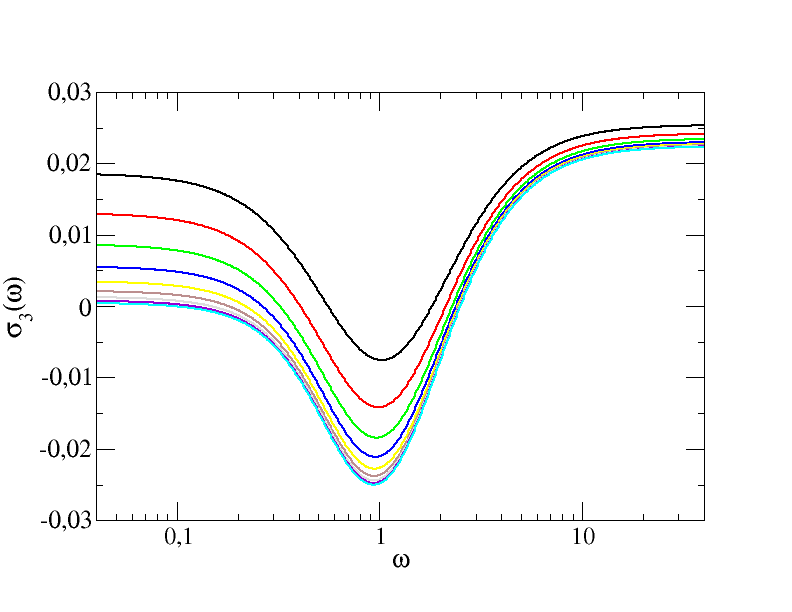}
\includegraphics[width=0.45\linewidth]{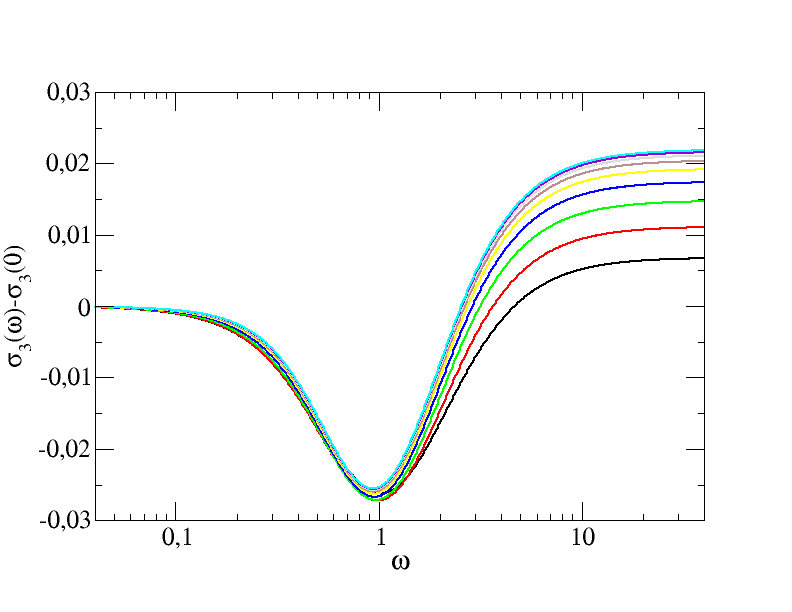}
\caption{Left: $\sigma_3(\omega)$ for fixed $V_2=1$ and for
different $V_3 \in [2,6]$, increasing from top to bottom in steps
of 0.5. Right: The same as before but with a shifted y-axis.}
\label{shift}
\end{figure}

One may wonder whether the occurrence of the minimum in
$\sigma_3(\omega)$ can be related to forward-backward motion in a
local double-well potential {, which would not contribute to an
overall conductivity}. If this were the case, the limit $V_3
\rightarrow \infty$ would already contain the relevant information
about this frequency regime. In order to elucidate this aspect {,}
we have calculated $\sigma_3(\omega)$ for a fixed value of $V_2$
and {various values} of $V_3$. In order to stay in the
experimentally relevant regime II we have chosen $V_2 = 1$. The
 {resulting graphs} are shown in Fig.\ref{shift} (left).
Interestingly, the
high-frequency plateau is nearly unchanged by $V_3$ and thus 
indeed just reflects the properties of the local DWP, governed by 
the asymmetry $V_2$. This can be directly read off from the
analytical expression Eq.\ref{s3i}. However, also the frequency
regime left of the minimum $\omega \approx 1$ displays some
specific properties. If one subtracts {$\sigma_3(0)$,} the
dc-value of the nonlinear conductivity{,} one finds a very similar
behavior for different values of $V_3$ in this low-frequency
regime $\omega \le 1$; see Fig.\ref{shift} (right). Thus, in this
frequency regime the nonlinear conductivity is basically a sum of
the dc-value and the contribution of the double-well potential,
described by {$\sigma_3(\w, V_3 = \infty)$}.  {It} turns out {,
though,} that this simple superposition principle becomes worse in
regime I or for $V_2 \approx 0$.

\begin{figure}
\centering
\includegraphics[width=0.45\linewidth]{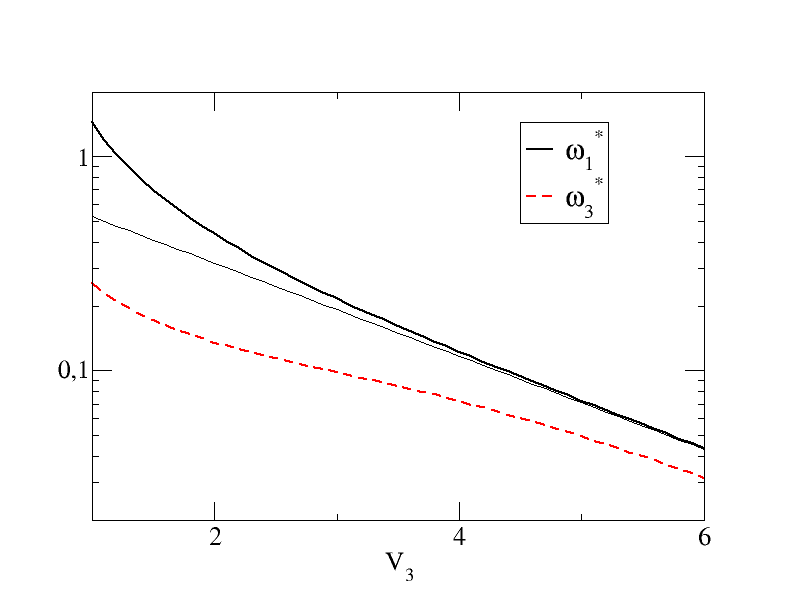}
\caption{The characteristic frequencies $\omega_1^*$ and
$\omega_3^*$ as a function of $V_3$ for fixed $V_2=1$. The thin
solid line corresponds to the approximation Eq.\ref{o1approx}.}
\label{omegastern}
\end{figure}

For a closer understanding of the nonlinear conductivity it may be
of interest to compare its characteristic frequencies with those
of the linear conductivity. We start by comparing the two
onset-frequencies $\omega_1^*$ and $\omega_3^*$ which characterize
the frequencies from which on deviations from the dc-regime are
relevant, i.e. where the conductivity increases or decreases,
respectively. Here we consider an increase and decrease of 10\%,
respectively, i.e. $\sigma_1(\omega_1^*)/\sigma_1(\omega=0)= 1.1$
and $\sigma_3(\omega_3^*)/\sigma_3(\omega=0)= 0.9$. The
qualitative behavior does not depends on this specific choice of
10\%.  $\omega_1^*$ can be calculated analytically from
Eq.\ref{sigmaall}. One obtains
\begin{equation}
\omega_1^* = \gamma_+ \frac{\sqrt{0.4 \Gamma_3}}{1-\Gamma_3}
\end{equation}
which for large $V_3$ (i.e. $\Gamma_3 \ll 1$) can be approximated
as
\begin{equation}
\omega_1^* \approx (1+\Gamma_2) \sqrt{0.4 \Gamma_3} \propto
\exp(-V_3/2) \label{o1approx}
\end{equation}

For the numerical analysis we restrict ourselves again to $V_2=1$
in order to avoid the regime III for better comparison with the
experimental situation. We have varied $V_3$ over a broad range,
encompassing the regimes I and II. As shown in
Fig.\ref{omegastern} there exists a significant dependence of
$\omega_1^*$ on $V_3$. The approximation Eq.\ref{o1approx}, i.e.
the scaling with $\exp(-V_3/2)$, roughly works for $V_3 \ge 3$.

Due to the complexity of $\sigma_3(\omega)$ one cannot write down
a simple analytical expression for $\omega_3^*$. Therefore we
restrict ourselves to the numerical determination of $\omega_3^*$;
see Fig.\ref{omegastern}. Two important observations can be made:
(1) The $V_3$-dependences of $\omega_1^*$ and $\omega_3^*$ are
identical for large $V_3$ (regime II). Both frequencies only
differ by a constant factor close to unity. (2) For small $V_3$,
i.e. in regime I, one observes $\omega_3^* \ll \omega_1^*$. We
would like to stress that this is fully compatible with the
experimental observations as reviewed in the Introduction
(inorganic ion conductors: regime II; ionic liquids: regime I).

\begin{figure}
\centering
\includegraphics[width=0.45\linewidth]{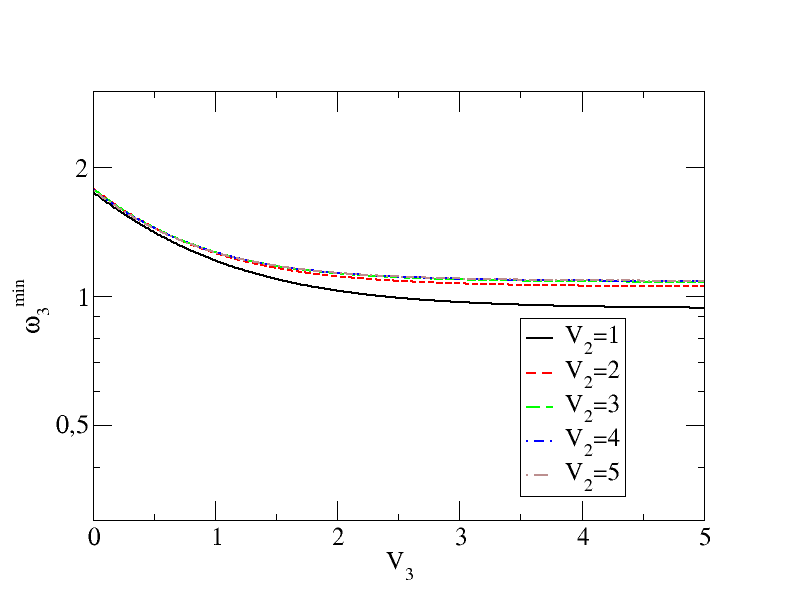}
\caption{The characteristic frequency $\omega_3^{min}$ as function
of $V_2$ and $V_3$.} \label{omega3}
\end{figure}

Another characteristic feature of $\sigma_3(\omega)$ is its
minimum, occurring at a frequency $\omega_3^{min}$. The data in
Fig.\ref{anacomp} already suggest that $\omega_3^{min}$ has no
strong dependence on $V_3$. We have analyzed its dependence on
$V_2$ and $V_3$. In Fig.\ref{omega3} we show more systematically,
how $\omega_3^{min}$  depends on $V_2$ and $V_3$. Obviously, the
dependence is very minor; less than a factor of two if taking into
account all possible parameter pairs ($V_2 \ge 1,V_3$).

This has several interesting implications: (1) Whereas variation
of $V_2$ and $V_3$ does not modify $\omega_3^{min}$ significantly,
this is not true for variation of $V_1$. As discussed above
consideration of the $V_1$-dependence gives rise to a trivial
factor $\exp(-V_1)$ for all rates and all characteristic
frequencies such as $\omega_3^{min}$. This exponential dependence
on $V_1$ thus has a significant impact on the frequency scales of
the conductivity. Thus, the minimum frequency of the nonlinear
conductivity is mainly sensitive to the typical barrier height
between adjacent minima rather than the asymmetry or the barrier
disorder. (2) Comparison of the linear and the nonlinear
conductivity in Fig.\ref{anacomp} leads to the assumption that the
appropriate counterpart to $\omega_3^{min}$ is the frequency where
the slope of $\sigma_1(\omega)$ is maximum in the
double-logarithmic-representation. We denote this frequency by
$\omega_1^{max}$. Starting from Eq.\ref{sigmaall} it can be
calculated analytically. One obtains
\begin{equation}
\omega_1^{max} = \gamma_+ \left ( \frac{(1+\Gamma_3)^2}{4
\Gamma_3}\right )^{1/4}= (1+\Gamma_2)(1+\Gamma_3)\left (
\frac{(1+\Gamma_3)^2}{4 \Gamma_3}\right )^{1/4}.
\end{equation}
Since the dependence on $\Gamma_2$ is only via the factor $(1 +
\Gamma_2)$, the dependence on $\Gamma_2$ vanishes for large $V_2$
in agreement with $\omega_3^{min}$. In contrast to
$\omega_3^{min}$ the frequency $\omega_1^{max}$ exponentially
depends on $V_3$ for large values of $V_3$ via $\exp(-V_3/4)$.
However, due to the factor $(1/4)$ in the exponent, the $V_3$
dependence is significantly weaker than for $\omega_1^* \propto
\exp(-V_3/2)$. Thus, if at all, the minimum of the nonlinear
conductivity is related to the region of maximum slope of the
linear conductivity  as evaluated in a double-logarithmic
representation.

\section{Conclusion}

The analyzed 1D hopping model may be considered as a minimum
model, which captures the non-trivial frequency dependence of the
nonlinear conductivity. Despite its simplicity, the algebraic
calculations are quite lengthy and some results required the help
of an algebra software. A key result was the presence of a minimum
of the nonlinear conductivity for all parameters except for the
trivial ordered case. The scaling properties of the different
characteristic frequencies of the linear and nonlinear response
allow one to see how the disorder influences the shape of the
frequency-dependent conductivities. In  comparison with the
experimental results on inorganic ion conductors and ionic liquids
it seems that the disorder effects are more pronounced in the
first case.

Of course, this model   mimics the true experimental system only
in a very simple way. First, one has to consider that in reality
one has a 3D rather than a 1D energy landscape. For example it
turns out that for a random barrier landscape, here corresponding
to $V_2=0$ one has $A(\omega=0)=1$ in the 1D system as discussed
above but $A(\omega=0)<0$ for the 3D system \cite{Roling2002}.
Second, in reality a disordered system is likely to be described
by a distribution of energy parameters rather than by well-defined
values. Of course, a natural next step would be to average the
present results over appropriately chosen parameter distributions
as performed, e.g., in \cite{Roling2002}.

Despite these limitations one may expect that important general features of the model may also hold for more complex systems. In particular the observation that the frequency dependence close to the minimum of the nonlinear conductivity in the experimentally relevant regime is to a large extent related to the dynamics in a local DWP may justify the use of the present model for a qualitative understanding of experimental systems.

\section*{Acknowledgements}

We gratefully acknowledge the support by the DFG (FOR 1394) as
well as very helpful discussions with M. L\"owe.


\end{document}